# Phonon-assisted formation of an itinerant electronic density wave

*Jiaruo Li[1], Oleg Yu. Gorobtsov[1], Sheena K. K. Patel[2], Nelson Hua[3], Benjamin Gregory[1], Anatoly G. Shabalin[3], Stjepan Hrkac[3], James Wingert[3], Devin Cela[3], James M. Glownia[5], Matthieu Chollet[5], Diling Zhu[5], Rajasekhar Medapalli[2,4], Eric E. Fullerton[2], Oleg G. Shpyrko[3], and Andrej Singer[1]*

[1]*Materials Science and Engineering Department, Cornell University, Ithaca, NY 14853, USA*
[2]*Center for Memory and Recording Research, University of California, La Jolla, CA 92093, USA*
[3]*Department of Physics, University of California, San Diego, La Jolla, California, 92093, USA*
[4]*Department of Physics, School of Sciences, National Institute of Technology, Andhra Pradesh 534102, India*
[5]*SLAC National Accelerator Laboratory, Menlo Park, CA 94025 USA*

**Electronic instabilities drive ordering transitions in condensed matter. Despite many advances in the microscopic understanding of the ordered states, a more nuanced and profound question often remains unanswered: how do the collective excitations influence the electronic order formation? Here, we experimentally show that a phonon affects the spin density wave (SDW) formation after an SDW-quench by femtosecond laser pulses. In a thin film, the temperature-dependent SDW period is quantized, allowing us to track the out-of-equilibrium formation path of the SDW precisely. By exploiting its persistent coupling to the lattice, we probe the SDW through the transient lattice distortion, measured by femtosecond X-ray diffraction. We find that within 500 femtoseconds after a complete quench, the SDW forms with the low-temperature period, directly bypassing a thermal state with the high-temperature period. We argue that a momentum-matched phonon launched by the quench changes the formation path of the SDW through the dynamic pinning of the order parameter.**

Peierls proposed a mechanism for the formation of charge density waves (CDW)[1]: a spin-less one-dimensional electron gas described by the Fröhlich-Hamiltonian is intrinsically unstable to the doubling of the unit cell. The spectrum of excitations develops an energy gap, and a CDW forms if the energy-gain from the electronic instability is larger than the strain-energy of the coexisting lattice distortion[2] (see Fig. 1(a)). A one-dimensional electron gas with spins described by the Hubbard-Hamiltonian develops a spin density wave (SDW), whose existence requires no lattice distortion[3]. In multidimensional systems, nesting between planar sections of the Fermi surface explains the presence of density waves[4,5]. Nevertheless, their explanation remains incomplete, especially the collective excitations' role during the density wave formation remains elusive. The ionic lattice distortions are necessary for the existence of CDWs[6]. Yet quasi-static measurements in chromium (Cr) – a prototypical system hosting a classical SDW, CDW[7,8], a pseudogap[9], and a quantum critical point[10] – assign the dominant role in electronic order formation to the itinerant instability with little impact from the ionic lattice[11].

Laser-induced quench and subsequent recovery of density waves offer a unique opportunity for studying the mechanisms of order formation in correlated systems[12-14]. A short laser pulse causes a spike in the electron temperature through the photon-electron interaction, which leads to density wave melting within 100 fs[12]. Because the electron gas's heat capacity is typically much smaller than that of the ionic lattice, the electronic subsystem cools via the electron-phonon interaction[15], and the density wave recovers within about one picosecond[14,15]. Often a description using the temperatures of different degrees of freedom captures the main features of the transition[14]. Here, we study the femtosecond quench and subsequent recovery of the SDW in Cr mediated by the generation and damping of an acoustic phonon (see Fig. 1a). We find the phenomenological Landau-model insufficient in describing the experimental observations. After a full quench, the SDW bypasses a state predicted by the model and observed experimentally in equilibrium and directly transitions into the low-temperature state. We hypothesize that the momentum-matched acoustic phonon, launched by the laser quench, guides the system along a non-equilibrium path in the free-energy diagram. Measurements with two precisely timed pump-pulses support our hypothesis of phonon-assisted SDW formation.

In a 28 nm thick epitaxial Cr film held at 115 K, the SDW ordering wavevector is perpendicular to the film surface (001). The periodic lattice distortion (PLD) – the second harmonic of the SDW[7] – is incommensurate with the lattice and generates two diffuse X-ray scattering peaks around the (002) Bragg peak at a momentum transfer of $Q_{(002)} \pm Q_L$. The interference between the diffuse scattering and Laue fringes results in coherent satellite-peaks (see Supplementary Material Fig. 1S). We excite the system with 30 fs optical laser pulses (λ=800 nm, see Fig. 1a) and monitor the satellite peak's intensity at $Q_{(002)}-Q_L$ with 10 fs X-ray laser pulses ($\lambda$=1.4 Å) [14,16,17] (see Fig. 1b and Methods). The time-dependent intensity consists of two components, an oscillation and an asymmetric distortion that is only present for time delays smaller than one ps. (see Fig. 1b).

We analyze the time-resolved X-ray data by splitting it into two intervals (see Fig. 1(b)). In interval II (1ps<t<5ps, blue shaded), we model the satellite-peak intensity as a damped harmonic oscillator

$$I(Q_{002} - Q_L, t) = P_F + P_A \cdot \cos\left(\frac{2\pi[t-t_0]}{t_w}\right) \cdot \exp(-[t - t_0]/t_D),$$

where the fitting parameter $P_F$ describes the baseline around which the intensity oscillates, $P_A$ is the amplitude at zero delay $t_0$, $t_W$ is the frequency of the oscillation, and $t_D$ is its damping time. To extract the non-oscillatory, asymmetric component of the PLD, we extrapolate the fit to shorter timescales (interval I: 0 ps<t<1ps) and subtract it from the data. The residual is the baseline around which the satellite peak harmonically oscillates (see Fig. 1b, yellow line).

The coherent satellite-peak intensity is proportional to the maximum displacement of the PLD[16], allowing us to quantitatively extract the dynamic lattice distortion directly from the X-ray data. The harmonic oscillation satellite-peak intensity is consistent with an amplitude mode of the PLD[6] – equivalent to a damped phonon. The laser quench launches a longitudinal acoustic phonon with a wave vector, $Q_L$, of the pre-existing static PLD – the laser pulse releases the previously "frozen-in" phonon[14]. By subtracting the harmonic oscillation from the total displacement, we determine the transient lattice displacement, $\tilde{A}_\psi(t)$, around which the lattice oscillates (see Fig. 2b and Methods). We interpret $\tilde{A}_\psi(t)$ as the transient SDW order parameter coupled to the measured lattice distortion through spin-lattice interactions. As expected, the order parameter displays a quench, followed by a recovery (see Fig. 1b).

Figure 2a shows the time-dependent intensity of the coherent satellite-peak as a function of the pump fluence. We first determine the phonon dynamics for each pump fluence by fitting the time-trace in interval II (1ps < t< 5ps). While the phonon's frequency (2.27 THz) and its damping time (3ps) are independent of the fluence, the phonon's maximum amplitude increases with fluence. At 3.5 mJ/cm² and above the laser pulse launches the phonon with the highest amplitude possible – approximately equal to the pre-quench amplitude of the periodic lattice distortion[14] (see Fig. 2S a). All data collected for t>1 ps is consistent with a damped harmonic oscillator, suggesting no phonon anharmonicity for all fluences.

The precise measurement of the oscillatory lattice distortion in the time domain allows us to extract the subtle non-oscillatory lattice distortion that couples to the transient electronic order. For all measured fluences, this transient order parameter, $\tilde{A}_\psi(t)$, displays a quench followed by a recovery (see Fig. 2b). The amount by which the transient order reduces, $\Delta\tilde{A}_\psi$, grows monotonically with the fluence before it saturates at a fluence of 3.5 mJ/cm² (see Fig. 2c). The saturation signifies the critical fluence beyond which we observe a full quench of the order parameter, likely because the electronic temperature rises above $T_N$ at this critical fluence[14]. A complete quench also yields a phonon with the maximum amplitude (see Fig. 2c and Fig. 2Sa). While the recovery is monotonic for lower fluences, our data displays a "shoulder" for high fluencies. A possible explanation for the shoulder is the phonon frequency dependence on charge carrier density, which is different after the optical quench[18]. An alternative reason is the discontinuity of the SDW recovery when the carrier temperature falls below $T_N$[19].

The order recovery time increases approximately linearly in the whole fluence range of 0-5.5 mJ/cm² (see Fig. 2d). Yet the observed laser-induced dynamics of the lattice in our experiment remain exceptionally fast; the order vanishes within 100 fs and recovers within 500 fs, even at the highest measured fluences. The recovery is a few times faster than in other density-wave-systems[20,21]. Nicholson[19] directly measured the SDW gap in Cr using time-resolved angle-resolved photoemission spectroscopy (tr-ARPES). Strikingly, our X-ray diffraction data and tr-ARPES show nearly identical time dependence, indicating persistent spin-lattice coupling out-of-equilibrium. It is this persistent coupling that we exploit to relate the SDW to the transient lattice distortion and to measure femtosecond SDW dynamics with non-resonant X-ray diffraction.

In the thin film studied here, the interfaces restrict the temperature-dependent SDW period in Cr[7] to two distinct values[22]. During quasi-static cooling, the SDW forms at $T_N$=290K with the wavevector $Q_H$ (7.5 PLD-periods across the film). At 230 K, the wavevector changes its wavevector to $Q_L$ (8.5 PLD-periods across the film) through an abrupt phase transition[22]. We correctly reproduce the quasi-static behavior using a phenomenological Landau model (see Fig.3a,b, and Methods)[10,11,13,23-25]. The superb momentum resolution of X-ray diffraction allows us to track the momentum transfer of the SDW during its formation out-of-equilibrium. Surprisingly, after a full laser quench, we observe no satellite peak at $Q_{(002)}-Q_H$: the SDW immediately forms with the low-temperature wavevector $Q_L$ (see Fig. 3b). The system directly bypasses a state expected from the free-energy diagram (compare gray and red lines in Fig. 3b,c). Additionally, the order recovers within one picosecond, suggesting the absence of an abrupt phase transition, which typically occurs on a significantly longer timescale (~100ps)[26,27].

In equilibrium, the SDW forms in the presence of Bose-Einstein distributed phonons at $T_N$. Out of equilibrium, the SDW forms at a lattice temperature lower than $T_N$ and in the presence of a momentum-matched phonon at $Q_L$ (see Fig. 3d). We hypothesize that the non-equilibrium phonon distribution dynamically pins the SDW, thus modifying its formation pathway in the free-energy landscape. We test our hypothesis of the phonon-assisted SDW-formation by exciting the system with two sequential laser pulses. Each laser pulse launches a phonon, and the delay between the laser pulses modifies the phase shift between the two phonons and the amplitude of their harmonic superposition. Varying the laser pulse delay by only 200 fs changes the superposition's amplitude from 0 to 1.6 times the amplitude of the pre-quench PLD (see Fig. 4, inset)[28]. When the laser pulse delay is larger than the SDW recovery time, we expect the second SDW-quench to be independent of the delay. Yet after the second quench, the SDW recovers faster for a larger phonon superposition amplitude, supporting our hypothesis of the phonon-assisted SDW formation (see Fig. 4 and Fig. 3S). In conclusion, our results show how a coherent ionic lattice vibration markedly modifies the pathway and the timescale of an electronic phase transformation, highlighting the opportunity for using collective excitations to guide electronic instabilities. Furthermore, we demonstrate how persistent spin-lattice coupling and non-resonant diffraction at X-ray lasers enable momentum-resolved interrogation of the transient electronic order in the presence of collective excitations.

**Acknowledgments:**

We thank Jerry Hastings for the discussions. The work was supported by the U.S. Department of Energy, Office of Science, Office of Basic Energy Sciences, under Contracts No. DE-SC0019414 (ultrafast X-ray data analysis and interpretation J.L., O.Y.G., B.G., A. S.), DE-SC0001805 (ultrafast X-ray scattering experiments, A.S., A.G.S., N.H., S.H., J.W., D.C., O.G.S.) and No. DE-SC0003678 (thin film synthesis and characterization, ultrafast

X-scattering, S. K. K. P., R. M., and E.E.F.). Use of the Linac Coherent Light Source (LCLS), SLAC National Accelerator Laboratory, is supported by the U.S. Department of Energy, Office of Science, Office of Basic Energy Sciences under Contract No. DE-AC02-76SF00515.

**Methods**

**Pump-Probe X-ray measurements**

The Cr film studied in this work was deposited onto a single-crystal MgO (001) substrate using DC magnetron sputtering at a substrate temperature of 500° C and annealed for an hour at 800° C. The growth process was optimized to yield both a smooth surface and good crystal quality of the sample. The film thickness was determined to be 28 nm by X-ray reflectivity. The Néel temperature of a thin film is 290K (see also refs. [14,17]). The pump-probe experiments were carried out at the XPP instrument of the LCLS with an X-ray photon energy of 8.9 keV, selected by the (111) diffraction of a diamond crystal. X-ray diffraction around the (002) Bragg peak ($2\theta = 60°$) from each pulse was recorded by an area detector (CS140k) with a repetition rate of 120 Hz. Due to the mosaic spread of the crystal in the film plane, several Laue oscillations appear on the area detector simultaneously. About 100 pulses were recorded for each time delay (50 fs steps in the time traces). For each time delay, the intensity was dark noise corrected and normalized by the intensity measured in the region of the area detector where Laue oscillations were absent. The sample was excited by optical (800 nm, 30-fs), p-polarized laser pulses propagating nearly collinear with the X-ray pulses. The temporal resolution at the sample plane was estimated to be ~80 fs. The spot sizes (full width at half maximum) of the optical and X-ray pulses were 0.46 mm (H) x 0.56 mm (V) and 0.2 mm (H) x 0.2 mm (V), respectively.

**Determining the order parameter from X-ray data**

We assume that the laser quench launches the phonon and that the order parameter's recovery has no impact on phonon amplitude and frequency. The uncertainty of our approximation is about 10 % of the order parameter because the maximum slope of the recovery is about three times smaller than the maximum slope of the phonon oscillation (proportional to the velocity of the oscillating atoms), adding about 10% to the phonon energy (proportional to the atomic displacement). We chose the robust regression fit with the least absolute residual method to minimize the outlier's influence and choose the upper and lower limits according to direct observation on the data shape. We subtract the extrapolated fit from the data, then add the PLD value at 4 ps to the residual. After 4 ps, the PLD represents an order parameter recovered to its new equilibrium value (slightly heated system). For fluences at around 11 mJ/cm$^2$ this amplitude vanishes[14], indicating the absence of the order parameter recovery because the system remains above the Néel temperature ($T_N$) (see Fig. 2S b), in agreement with mean-field theory[3,29]. The subsequent heat transport from the film into the substrate occurs in ~0.3 ns[16]. The laser fluence was determined by comparing the data directly with the published data on the exactly the same films, where the fluence was determined by using the Bragg peak position as a thermometer[14].

**Landau-theory modelling**

We use the Landau theory developed by McMillan for describing density waves[30]. It describes how density wave systems evolve microscopically and agrees with experiments in multiple CDW systems and SDW in Cr under quasi-static and pressure-driven conditions [10,11]. We adapt the free-energy density to a thin film Cr introducing terms similar to [10] as

$$F(x) = a\psi^2 + b\psi^4 + c\xi|\nabla\psi|^2 + e\psi^2|\nabla\phi - q|^2 + f\psi^4\cos(2\phi - Gx)$$

where $\psi e^{i\phi}$ is the complex SDW order parameter with the real amplitude $\psi$ (proportional to $\tilde{A}_\psi$) and phase $\phi$. Here, $q$ is the natural nesting SDW wavevector given by the electronic instability, $G$ the wave vector enforced by boundary conditions, and $a, b, c, d, e, f$ the coefficients. The first two terms are the second-order free-energy expansion of Landau theory. We assume $a = a_0(T - T_N)$, where $T_N$ is the Néel temperature, and $b, c, d, e, f$ are temperature independent. The third term is electronic energy from finite spatial correlation length. We neglect the third term by assuming $\psi$ is uniform across the thin film, i.e., $\nabla\psi = 0$. The fourth term is the energy cost of changing SDW wavevector from the natural wavevector.

We use the fifth term to model the energy of the lock-in to the boundary conditions (possibly by impurities at the interfaces [31]), which restrict the available wave vectors of the SDW in a thin film. Specifically, we do not use G as a reciprocal space vector[30], we use $G = Q_L$ and $G = Q_H$ to enforce the pinning by the interfaces. We replace the last term with $f_1\psi^4\cos(2\phi - 2Q_Hx) + f_2\psi^4\cos(2\phi - 2Q_Lx)$, $|f_2| > |f_1|$. This modification of McMillans formalism allows us to reproduce the experimentally observed temperature dependence: the SDW forms with $Q_H$ and subsequently transitions to $Q_L$.

We integrate free-energy density $F$ over film thickness $L$ by using $\phi = Qx$ and find

$$F(\psi, Q) = L(a\psi^2 + b\psi^4 + c\xi|\nabla\psi|^2 + e\psi^2|Q-q|^2) + f_1|\psi|^4 \frac{\text{sinc}(2L(Q-Q_L))}{2(Q-Q_L)} + f_2|\psi|^4 \frac{\text{sinc}(2L(Q-Q_H))}{2(Q-Q_H)}.$$

The above equation is used to calculate the free energy surface shown in Figs.3a,b. For a given temperature (value of $a$) and each value of the order parameter $\psi$, we minimize $F(\psi, Q)$ as a function of $Q$ to determine the path of Q during the formation of the density-wave. Figure 3a shows this minimized $F(\psi, Q_m)$, which is dominated by the competition between the first two Landau terms. Similarly, for a given temperature and each value of the wavevector $Q$, we minimize $F(\psi, Q)$ as a function of $\psi$. Figure 3b shows this minimized $F(\psi, Q_m) - F(\psi_m, Q_m)$. This function displays the competition between the intrinsic wavevector q, and the two wavevectors restricted by the film geometry. Because only two values $Q_H$ and $Q_L$ are possible, switching between the two requires overcoming an energy barrier.

**Figures**

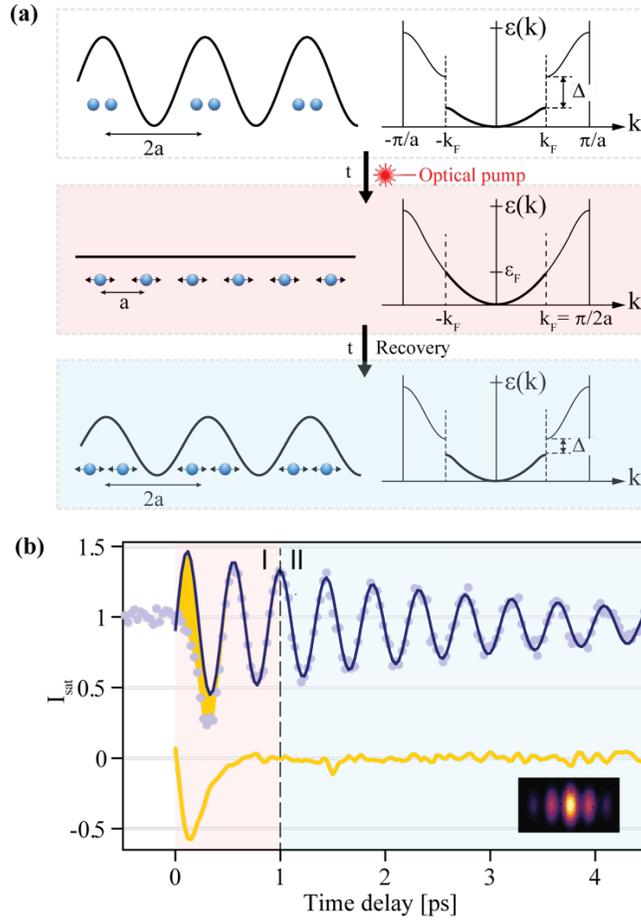

**Figure 1: Time resolved measurement of the electronic order formation in presence of an acoustic phonon. (a)** The laser pulse quenches the SDW, which is coupled to the PLD (solid line on the left), and launches a longitudinal acoustic phonon (atomic positions (blue spheres) and their momentum (black arrows)). Subsequently, the SDW bandgap (schematic energy diagram on the right) reopens and the coupled SDW-PLD state reforms in presence of the phonon. **(b)** Typical time-resolved X-ray diffraction data of the coherent satellite peak (purple dots) for a pump fluence of 2mJ/cm². The data is fitted (purple solid line) with a damped harmonic oscillator in region II (blue shaded area), is extrapolated to region I (red shaded area), and subtracted from X-ray diffraction data. The residual is shown in orange and is proportional to the transient PLD amplitude coupled to the SDW through spin-lattice interaction. The inset shows a typical diffraction data as measured on the two-dimensional detector for a single time delay. The coherent satellite peak is in the center at $Q_{(002)}$-$Q_L$=4.17Å$^{-1}$, the image spans 1.35 nm$^{-1}$ horizontally, and the intensity drop towards the sides of the detector is due to the Ewald sphere dissecting the Bragg rod under an angle of 30 degrees.

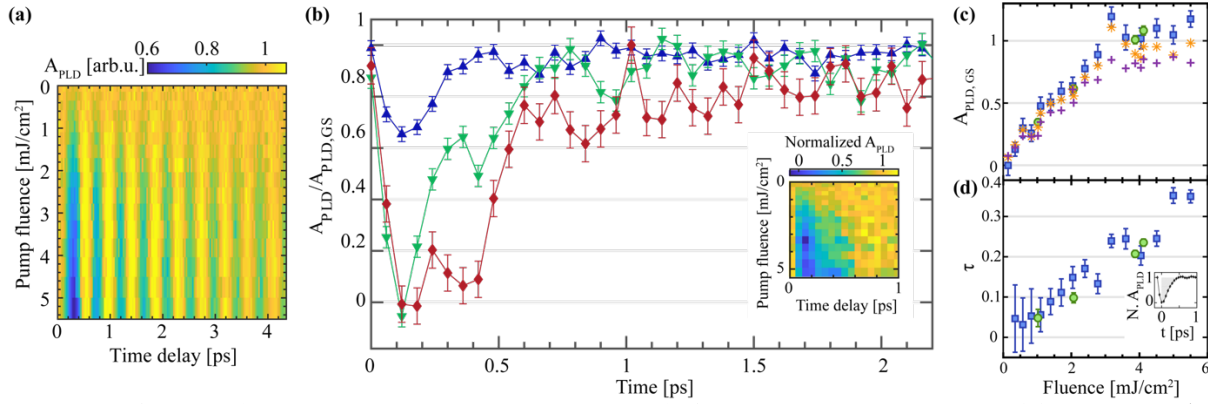

**Figure 2: Fluence dependent phonon dynamics and electronic order recovery. (a)** Time-resolved X-ray satellite-peak intensity as a function of the laser pump fluence normalized by the pre-quench PLD amplitude. **(b)** The amplitude of the transient order parameter, $\tilde{A}_\psi(t)$, normalized by the amplitude in the ground state and shown for three different fluences: 1 mJ/cm² (blue), 3 mJ/cm² (green) and 5 mJ/cm² (red). The inset shows the transient PLD amplitude for all fluences measured in **(a)**. **(c)** The SDW quench amount $\Delta\tilde{A}_\psi = \tilde{A}_\psi(0\,ps) - \tilde{A}_\psi(0.1\,ps)$ (blue squares: data from **(a)** and green circles: higher statistics data) and the recovery amount (orange stars) determined from the transient PLD amplitude shown in **(b)**. The amplitude of the phonon is also shown (magenta crosses). **(d)** The recovery time determined as the area indicated in the inset, where $\tilde{A}_\psi$ was rescaled to $\tilde{A}_\psi(0.1ps) = 0$ and $\tilde{A}_\psi(1\,ps) = 1$. The integral is taken from 0.1 ps to 1.1 ps. We choose the area because the exponential fit poorly reproduces the data at higher fluences. The uncertainty in **(b, c, d)** is estimated as the standard deviation of the data at time delays >1 ps, where we expect the residual between the harmonic oscillator fit and the date to be zero.

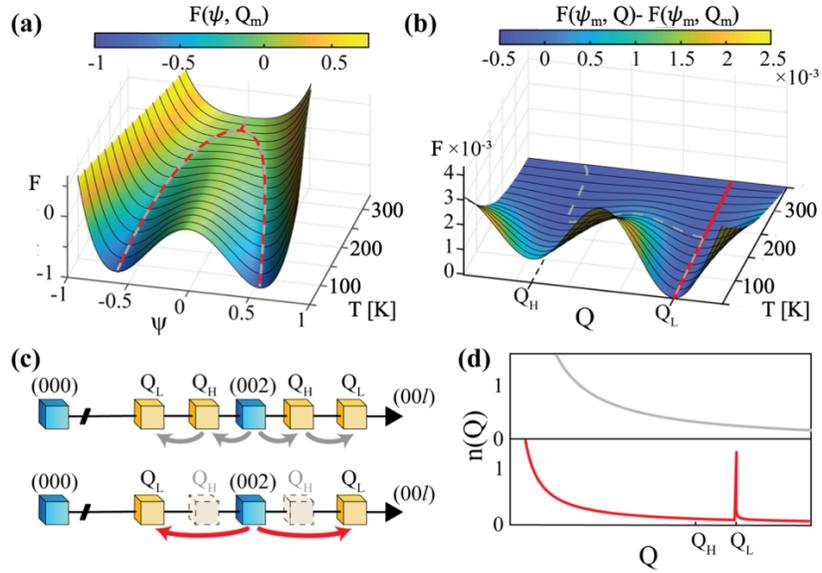

**Figure 3: Landau model of the electronic order parameter and its ordering wavevector. (a)** The Landau free energy, $F(\psi, Q_m)$, as a function of temperature and the order parameter $\psi$, shown for the momentum transfer $Q_m$ that minimizes the free energy. **(b)** The free energy $F(\psi_m, Q)$ as a function of temperature and momentum transfer Q, shown for the order parameter $\psi_m$ that minimizes the energy. In **(a)** and **(b)** the equilibrium (gray dashed line) and non-equilibrium (red solid line) paths during order formation are shown. **(c)** The wavevector dynamics revealed through measuring the satellite peak intensity and position: in equilibrium, the SDW forms with a smaller wavevector $Q_H$ and then transitions to $Q_L$ via an abrupt (first-order) phase transition (consistent with the energy barrier indicated by the gray dashed line in **(c)** Out-of-equilibrium, the order immediately forms with $Q_L$, bypassing the state at $Q_H$ (see red line in **(c)**). **(d)** A schematic phonon distribution during the formation of the SDW: in equilibrium the SDW forms at the Néel temperature (top, gray), and out-of-equilibrium the SDW forms at a lower lattice temperature and in presence of a momentum-matched acoustic phonon at $Q_L$ (red, bottom).

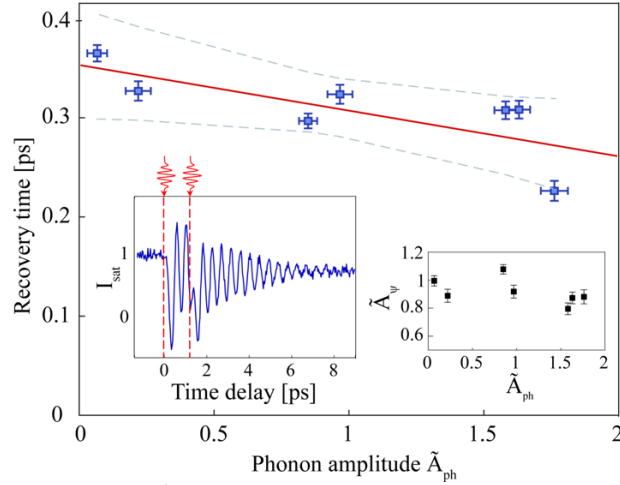

**Figure 4: Testing the hypothesis of the phonon-assisted electronic order formation by using two pump-pulses.** The density wave recovery time as a function of the phonon amplitude, both measured after the arrival of the second pump pulse. Each pair of values is determined from a fit to the data measured with a fixed time delay between the two pump pulses; the analysis procedure is identical to the analysis used for single pulse data, with taking the arrival of the second pulse as time zero. The red line shows a linear fit with $\tau(\tilde{A}_{ph}) = b\tilde{A}_{ph} + c$, with $b = -0.046$ and $c = 0.35$. The dashed gray lines indicate the prediction intervals of the fit. Left inset: Typical time-resolved X-ray data for two pump-pump pulses as indicated. The second pulse changes the amplitude of the phonon. Right inset: The quench after the second pulse as a function of the phonon amplitude: the second pulse induces the same quench independent of the phonon amplitude. All pump-pump delays are larger than 0.5 ps, ensuring the electron-lattice equilibration completes before the second pulse arrives. Both pump pulses have the same intensity, 1 mJ/cm², which only partially quenches the SDW.

# Supplementary Materials for

**Phonon-assisted formation of an itinerant electronic density wave**

Jiaruo Li[1], Oleg Yu. Gorobtsov[1], Sheena K. K. Patel[2], Nelson Hua[3], Benjamin Gregory[1], Anatoly G. Shabalin[3], Raj Medapalli[2], Stjepan Hrkac[3], James Wingert[3], Devin Cela[3], James M. Glownia[4], Matthieu Chollet[4], Diling Zhu[4], Eric E. Fullerton[2], Oleg G. Shpyrko[3], Andrej Singer[1]

[1]Materials Science and Engineering Department, Cornell University, Ithaca, NY 14853, USA
[2]Center for Memory and Recording Research, University of California, San Diego, La Jolla, California, 92093, USA
[3]Department of Physics, University of California, San Diego, La Jolla, California, 92093, USA
[6]SLAC National Accelerator Laboratory, Menlo Park, CA 94025 USA

**X-ray diffraction structure and behavior**

X-ray diffraction from a crystalline chromium film takes shape of a multitude of Bragg peaks with Laue fringes in the direction perpendicular to the film surface. In the experiment, we concentrate on the (002) peak, with (001) being an out-of-plane direction. Below the Neel temperature, when a SDW forms in the film and leads to a periodic lattice distortion (PLD), the PLD gives rise to satellite peaks (0 0 2-$\delta$) and (0 0 2+$\delta$). In a thin film, PLD is directed out-of-plane, and the satellite peaks will overlap with Laue fringes. More precisely, finite number of PLD periods in the film, limited also by boundary conditions on the film surface, means that satellite peaks will be "on top" of fringes corresponding to the number of PLD periods. Interference between the satellite peaks and the Laue fringes is *additive* at (0 0 2-$\delta$) and *destructive* at (0 0 2+$\delta$) (see Fig. S1, a). In experiment, we can only observe simultaneously the fringes lying on the Ewald sphere. By adjusting the sample and detector position, we position the maximum of the PLD fringe on the detector (Fig. S1, a). In the diffraction pattern (Fig. S1, b) the PLD fringe and the neighboring fringes cross-sected by the Ewald sphere are then clearly visible. The edge of the main Bragg peak also appears on the detector (Fig. S1, b, far left) and serves as a reference point.

After the photoexcitation, the excited acoustic phonon in the film with the same spatial period as the PLD leads to an oscillation in the PLD fringe intensity (Fig. S1, c). Thermal expansion of the lattice also moves the Bragg peak in the negative q direction after ~1 ps, however the total movement is <5% of the PLD fringe width for the 30 nm film. The total change in the fringe intensity can then used to track the changes in the lattice distortion.

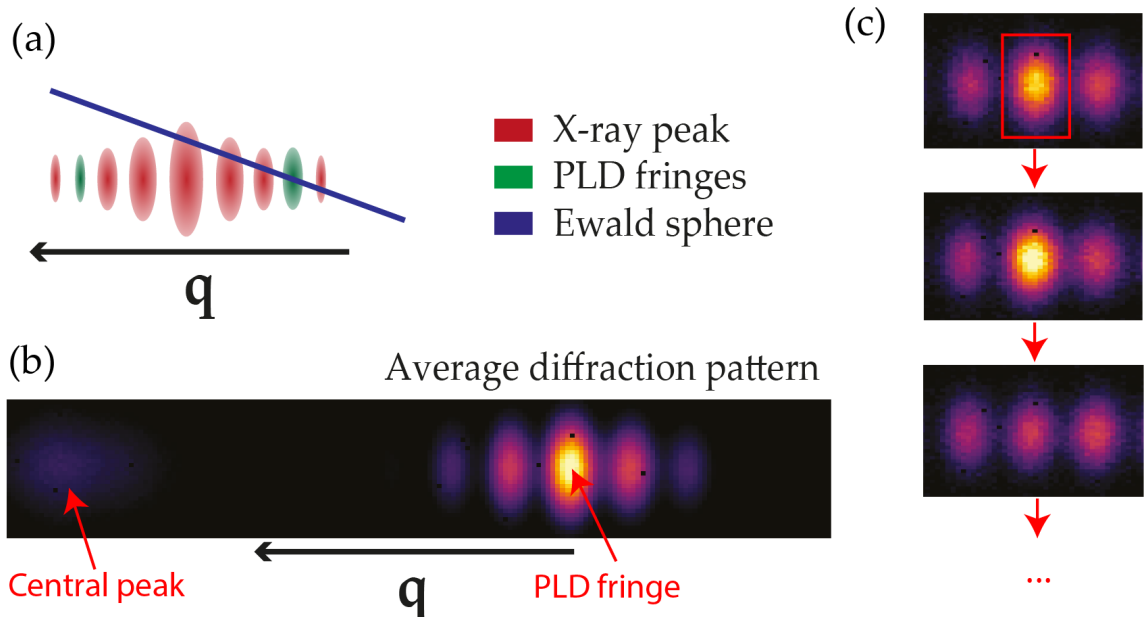

**Fig. 1S. X-ray diffraction structure. a,** X-ray scattering from periodic atomic displacement is detected on Laue fringes of the main peak from the crystalline film. **b,** Average static diffraction pattern observed in the experiment, with a main peak, PLD fringe and neighboring fringes visible. **c,** Oscillation of the PLD fringe after excitation.

**Fluence effects on the amplitude of the acoustic phonon and PLD after thermalization**

The amplitude of the acoustic phonon excited by the laser depends on both the static PLD and the energy introduced by the laser. If the electron temperature after the photoexcitation is below Neel temperature, the remaining PLD limits the amplitude of the phonon (Fig. 2S, a). If the SDW is fully melted, additional energy does not affect the amplitude of the excited phonon. However, the energy introduced can further affects the final level of the PLD after thermalization of the electron gas with the lattice (Fig. 2S, b), since the lattice temperature after thermalization can still be below the critical point. The dependence of the PLD on the laser fluence then follows the behavior of the order parameter as a function of temperature, since the temperature of the film after thermalization (with itself, not with the substrate) is essentially proportional to the laser energy. Compare (Fig. 2S, b) with for ex. [Hill1995]: the PLD decreases linearly with fluence, however the change in PLD is significantly faster closer to the critical point. At the fluence of ~11 mJ/cm$^2$, the lattice temperature reaches the critical temperature, and further increase in the laser intensity does not lead to changes in the final PLD. The final static PLD is non-existent due to the absence of the spin order.

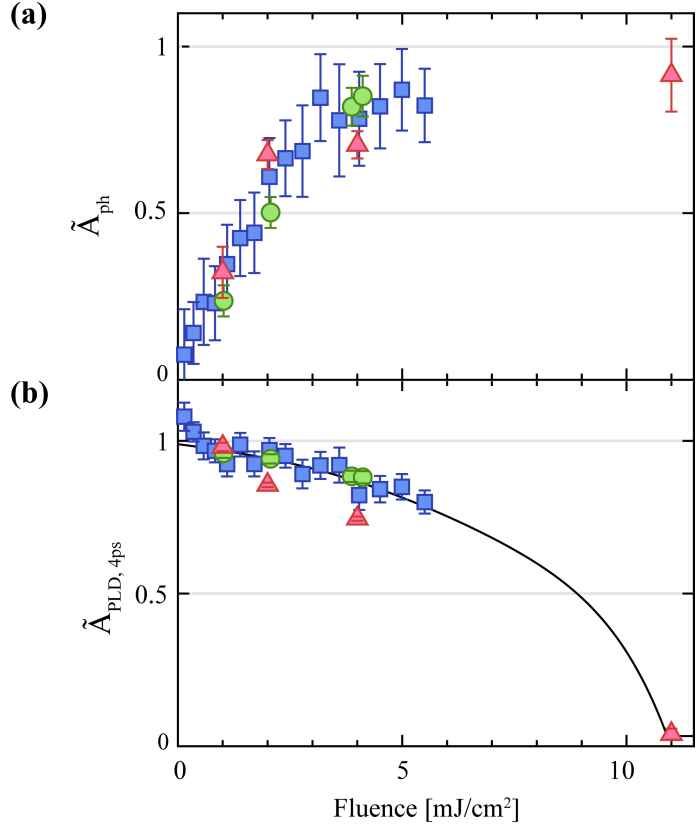

**Fig. 2S. X-ray diffraction structure. a,** Amplitude of the acoustic phonon at zero delay as a function of fluence. **b,** Static amplitude of the PLD after thermalization with the lattice, assuming the dampening of the acoustic phonon, as a function of fluence.

**Additional data, phonon-assisted SDW formation**

In addition to the data presented in Fig. 4 of the main paper, we measured additional datasets with varying delay between optical and x-ray pulses. The datasets themselves have been presented in [1]. We have performed the analysis of the recovery time on them in the same way as for Fig. 4. Due to the lower number of collected x-ray pulses for each delay, the experimental error is higher, however the amount of delays is larger. The results, presented in Fig. S3 a,b, show the same tendencies as Fig. 4, with the recovery time decreasing with the acoustic phonon amplitude and the SDW quench independent of the delay or the phonon amplitude.

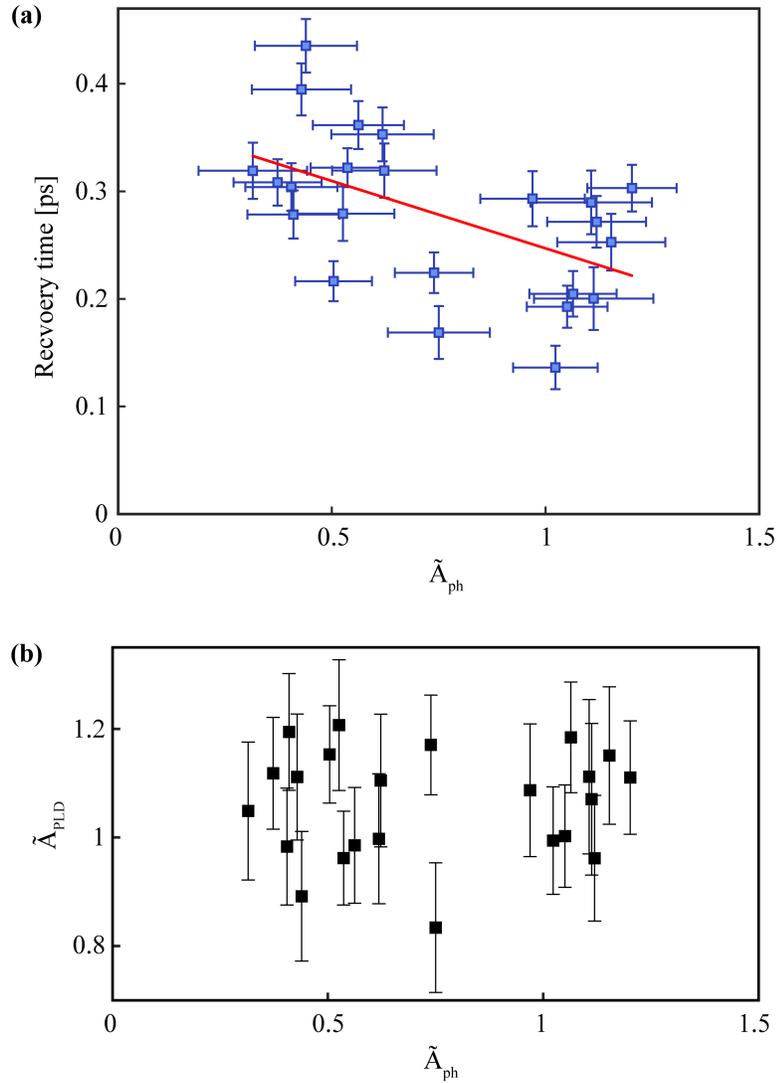

**Fig. 3S. Phonon-assisted SDW formation, additional data. a,** Recovery time after the second photoexcitation as a function of the amplitude of the excited acoustic phonon. **b,** Static amplitude of the PLD after thermalization with the lattice as a function of the amplitude of the excited acoustic phonon, after the second photoexcitation.

[1] Gorobtsov et al., Femtosecond control of phonon dynamics near a magnetic order critical point, in review 10.21203/rs.3.rs-40430/v1.